\documentclass[12pt]{article}         
\usepackage{amsmath,amssymb,dsfont}
\newcommand{\la}{{\lambda}}
\newcommand{\al}{{\alpha}}

\newcommand{\jmp}[3]{J. Math. Phys. {\bf #1}, #2, (#3)}
\newcommand{\Ha}[3]{{H^{(#1)}_{#2,#3}}}
\def\alg{{\mathfrak g}}

\def\be{\beta}
\def\back{\!\!\!\!\!\!}

\begin{document}
\begin{center}
{\large\bf Quantisation of Bending Flows}
\end{center}
\vspace{0.2truecm}
\begin{center}
{Gregorio Falqui \\
Dipartimento di Matematica e
  Applicazioni, Universit\`a di Milano--Bicocca,  via R. Cozzi 53, 20125
  Milano, Italy, gregorio.falqui@unimib.it\\
Fabio Musso\\
Dipartimento di Fisica, Universit\`a di
  Roma III, and INFN, Sezione di Roma III, via Vasca Navale, 84, 00146 Roma,
  Italy, musso@fis.uniroma3.it}
\end{center}
\vspace{0.1truecm}

\begin{abstract}\noindent
We briefly review the Kapovich-Millson notion of Bending flows as an
integrable system on the space of polygons in ${\bf R}^3$, its
connection with a specific Gaudin XXX system, as well as the
generalisation to $su(r), r>2$. 
Then we consider the quantisation problem of the set of Hamiltonians
pertaining to the problem, quite naturally called Bending
Hamiltonians, and prove that their commutativity is preserved at the
quantum level.

\end{abstract}
\section{Introduction: Classical Bending Flows and Gaudin Models}
\label{sect:1} Bending flows were introduced by Kapovich and Millson
(KM) in \cite{KM96}, as particular Hamiltonian integrable systems  on
the manifold of ``moduli'' of $n$-sided polygons ($n$--gons) in
$\mathbf{R}^3$. They can be briefly described as follows.

An $n$-gon in Euclidean $3$-space is specified by its $n$ vertices
$\{v_1,\ldots,v_n\}$, or, up to a translation, by its $n$  sides
$\{e_1,\ldots, e_n\}$, the lengths of the elements of the latter
being given by a suitable string $\mathbf{r}=\{r_1,\ldots, r_n\}$ of
non-negative numbers. The moduli space of such $n$-gons, ${\cal
M}_{n,\mathbf{r}}$, is obtained by factoring out the action of the
Euclidean group on the set of $n$-gons with preassigned
side--lengths.
It can be identified with the Hamiltonian
reduction of the Cartesian product $P_{n,\mathbf{r}}$ of $n$
two-spheres $\mathbf{S}^2_i$ of radius $r_i$ (endowed with the
standard symplectic two-form), with respect to the Hamiltonian
action of the group $SO(3)$, the closure condition of the $n$-gon
being translated in the fact that the value of the moment map to be
considered is the zero value. Thus the {\em moduli space\/} of
$n$-gons in ${\mathbf R}^3$, ${\cal M}_{n,\mathbf{r}}={\cal
P}_{n,\mathbf{r}}//_{SO(3)},$ is a $2n-6$ dimensional symplectic
manifold.

A completely integrable classical Hamiltonian system on ${\cal
M}_{n,\mathbf{r}}$ can be defined (\cite{KM96}) on this phase space
as follows. Fix one of the vertices of a polygon, say $v_1$, and
consider the $n-3$ diagonals $\{d_1,\ldots,d_{n-3}\}$ stemming from
$v_1$, and their squared lengths $h_\alpha=||d_\alpha||^2$. It can
be proven that the functions $h_\alpha$ are mutually in involution
and functionally independent on ${\cal M}_{n,\mathbf{r}}$, and thus
give rise to a Liouville integrable system. The name {\em Bending
flows} for these Hamiltonian systems comes from the fact that the
motion induced by the Hamiltonians $h_\al$ is a rotation (with
constant velocity) of the sub-polygon defined by the edges
$\{e_1,e_2,\ldots, e_{\al+1}, d_\al\}$ around the diagonal $d_\al$,
while the complementary sub-polygon $\{d_\al,e_{\al+2},\ldots,
e_{n}\}$ is left fixed in $\mathbf{R}^3$. In other words, ``half''of
the $n$--gon {\em bends} around the $\al$-th diagonal.

In \cite{FM03} a connection of these flows with (a particular case
of) XXX classical Gaudin model was described. Indeed, the KM moduli
space coincides with a subset of the phase space of such  $su(2)$
Gaudin magnet with $N=n-3$ sites, and the KM Hamiltonians $h_\al$
were shown to be the classical analogue of a set of mutually
commuting operators found by Ballesteros, Ragnisco and collaborators
(see, e.g., \cite{Balrag}) in their study of Hopf algebraic
properties of quantum Gaudin magnets.

The classical limits of these Hamiltonians are represented as
follows. It is known that the classical Gaudin system admits a Lax
representation, with Lax matrix
\begin{equation}
  \label{eq:2.2}
  L(\lambda)=\sum_{i=1}^N\frac{A_i}{\lambda-z_i},
\end{equation}
the $z_i$ being a set of numerical parameters, while the $A_i$ being
$su(2)$-valued matrices that encode the degrees of freedom of the
model. The Hamiltonians introduced by Ballesteros and Ragnisco, that
we herewith consider are given by
\begin{equation}
  \label{eq:2.3}
  J_\al=\frac12{\mathrm Tr}\left(\sum_{i=1}^{\al+1}A_i\right)^2, \quad
  \al=1,\ldots, N-1.
\end{equation}
Together with a component of the total spin these $J_a$'s provide a
set of mutually commuting integrals of the motion for the ``physical''
Gaudin Hamiltonian
\begin{equation}
  \label{eq:2.4}
  H_G=\frac12\sum_{i\neq j=1}^{N} {\rm Tr}(A_iA_j),
\end{equation}
that is, alternative to the ``standard'' one, namely,
the set of spectral invariants of the Lax matrix
(\ref{eq:2.2}), i.e.:
\begin{equation}
  \label{eq:2.5}
  H_\alpha={\rm Res}_{\lambda=z_\alpha} {\rm Tr} L(\lambda)^2.
\end{equation}
This identification suggests the opportunity to study, among others,
the following three problems:
\begin{description}\label{pl:1}
\item{A)} Generalise bending flows to the $su(n)$ case, and to
a general Lie algebra $\mathfrak{g}$.
\item{B)} Use the identification to provide ``new'' ways of solving the
classical counterparts of the Gaudin models encompassed by this framework.
\item{C)} Tackle the quantum case (with $\mathfrak g$-valued spins)
from this standpoint.
\end{description}
Points A and B above were described and substantially
solved in \cite{FM03} and \cite{FM04}. The main idea was to use tools from the
bihamiltonian theory of integrable systems. The basic point is that the phase
space of $\mathfrak g$-valued Gaudin models is, as it is well known, a
symplectic submanifold ({\em a leaf\/}) in the tensor product $({\mathfrak
  g^*})^{\otimes N}$ of $N$ copies of the dual of the Lie algebra
$\mathfrak g$, endowed with the standard Lie--Poisson brackets.
A suitable additional Poisson structure was defined on $({\mathfrak
  g^*})^{\otimes N}$ in \cite{FM03} with the following properties:\\
1) The Poisson brackets defined by this new structure are compatible
(in the Magri sense) with the Lie-Poisson brackets, so that they
define, on $({\mathfrak g^*})^{\otimes N}$, a {\em pencil\/} of
Poisson brackets.\\
2) The Lenard-Magri sequences defined by this pencil give rise, in
the ${\mathfrak g}=su(2)$ case, to the KM integrals (\ref{eq:2.3});
in the case of
  general simple $\mathfrak{g}$, the bihamiltonian iteration provides a set
  of (bi--involutive) integrals of the motion; they were called
  ({\em generalised}) Bending Hamiltonians. These Hamiltonians,
when complemented with a suitably chosen set of integrals associated with the global invariance under
  the Lie group $G=\exp(\mathfrak{g})$, insure the Liouville
integrability of the model.

An instrumental feature of this scheme is the following. The
integrals of the motion associated with point 2) of the above list,
can be obtained as follows. For $a=2,3, \ldots N$ one can introduce
$\alg$-valued matrices
\begin{equation}
  \label{eq:2.6}
  L_a(\la)=(\la-(a-2))A_\alpha+\sum_{i=1}^{a-1} A_i.
\end{equation}
These matrices are Lax matrices for the Hamiltonian flows associated with the
generalised Bending Hamiltonians, that is, they evolve isospectrally
along the generalised Bending flows; the ring of their spectral invariants
coincides with that of the generalised Bending Hamiltonians.

Furthermore, these Lax matrices provide, according to the so-called
Sklyanin magic recipe \cite{Sk95} and the bihamiltonian scheme of
Separation of Variables (see, e.g., \cite{FP03}), a set of
separation coordinates. It can be noticed \cite{FM03} that, in the case of
$\mathfrak{g}=sl(r)$, the solution to the HJ equations involves
integration of Abelian differentials on Riemann surfaces of genus
$g=(r-1)(r-2)/2$, irrespectively of the number $N$ of sites (while
the SoV scheme associated with the ``single'' Lax matrix
(\ref{eq:2.2}) involves a Riemann surface whose genus grows linearly
with $N$).

What is more important for the purpose of the present paper is that
they will be the basic objects for the quantisation of the Bending
Hamiltonians (point C of the problem list of the previous page), to
be discussed in the sequel.\newpage
\section{Quantisation of the bending Hamiltonians}\label{sect:2}
The problem of the quantisation of Gaudin Hamiltonians (with
$\alg$-valued spins) was discussed (see, e.g., \cite{FFR} and the
references quoted therein), via the quantisation of the spectral
invariants of the ``Sklyanin'' Lax matrix (\ref{eq:2.2}), and their
diagonalisation via Bethe Ansatz techniques. Here we will consider
the spectral invariants associated with the Lax matrices of the
Bending Flows(\ref{eq:2.6}).

As in the usual case, the matrices $A_i$ (and hence the Bending Lax
matrices $L_a$) will become operators in a suitable Hilbert space,
which, in the present paper, will be the $N$-th tensor product of a
faithful representation $\rho$ of a simple Lie algebra $\alg$. Their
quantisation is given by the following recipe:
\begin{eqnarray}
&& A_i= g_{\al \be} \rho(X^{\al}) y^{\be}_i \  \rightarrow
\ A^{(q)}_i= g_{\al \be} \rho(X^{\al}) X^{\be}_i
\label{Aqi} \\
&& L_a=\la A_{a} + \sum_{k=1}^{a-1}( A_k) \  \rightarrow \  L^{(q)}_a
=\la A^{(q)}_{a}
+
\sum_{k=1}^{a-1}( A^{(q)}_k)
\label{Lqi}
\end{eqnarray}
Here, $g_{\al \be}$ are the components of the inverse of the Cartan
matrix of $\alg$ and we shifted the spectral parameter appearing in
(\ref{eq:2.6}). In \cite{Balrag} it was already noticed that the
straightforward quantisation of the specific Hamiltonians
${\rm{Tr}}\big(( \sum_{k=1}^j A_k)^i\big) $ do commute (without any
quantum correction). However, the number of such Hamiltonians grows
linearly with the number $N$ of sites of the magnet, and so cannot
provide a complete set of commuting operators for generic
representations $\rho$ of $\alg$ whenever $\alg\neq su(2)$.

In this section we will extend this result to a larger set of
generalised bending Hamiltonians, using in the following the
$i-$coproducts in the universal enveloping algebra,
$\Delta^{(i)}:U(\alg) \longrightarrow U(\alg)^{ \otimes i}$:
\begin{equation}
\Delta^{(i)}(1)=1, \quad \Delta^{(i)}(X^\al)=\sum_{k=1}^i X^\al_k,
\>\> X^\al \in \alg,\quad \Delta^{(i)}(ab)= \Delta^{(i)}(a)
\Delta^{(i)}(b)\,.
\end{equation}
We will make use of the following identity:
\begin{equation}\label{qlemma}
\left[ {\rm{Tr}} \left( L_a^{(q)}(\la)^m \right) , \Delta^{(p)}(X^\al) \right]
=0 \quad  {\rm{for}} \ p>a
\quad X^\al \in \alg,
\end{equation}
whose proof goes as follows.

>From the fact that ${\rm{Tr}}(A_i^m)$ give us central elements of
$U(\alg)$ we know that, for $X^\al \in \alg$, and every index $i$,
it holds that $[ {\rm{Tr}}(A_i^m), X^\al_i]=0$. Defining:
\begin{equation}
g_{j_1 \dots j_m}:= g_{i_1 j_1} g_{i_2 j_2} \dots g_{i_m j_m}
{\rm{Tr}} \left( \rho(X^{i_1}) \dots \rho(X^{i_m}) \right)
\end{equation}
one can obtain the identity $\sum_{r=1}^m  g_{j_1 \dots
j_{r-1} s j_{r+1} \dots j_m} C^{sq}_{j_r}=0$ for every
$\,j_1,\dots,j_m$. Using this and the commutation relations
($\Delta^{(i)}$ is a Lie algebra homomorphism)
\[
\left[  X^{j}_{i+1} ,  \Delta^{(p)}(X^q) \right]=C^{jq}_s
X^{s}_{i+1},\qquad
\left[  \Delta^{(i)}(X^{j}) ,  \Delta^{(p)}(X^q) \right]=C^{jq}_s \Delta^{(i)}(X^{s})
\quad p>i\, ,
\]
we can conclude
\begin{eqnarray*}
&&  \back \left[ {\rm{Tr}}\left( (L_i^{(q)}(\la))^m \right), \Delta^{(p)}(X^q) \right]
=\left[ {\rm{Tr}}\left( \left( \la A_{i+1} + \Delta^{(i)}(A) \right)^m \right), \Delta^{(p)}(X^q) \right]=\\
&& \back = \left[ g_{j_1 \dots j_m} (\la X^{j_1}_{i+1}+ \Delta^{(i)}(X^{j_1}) \dots
(\la X^{j_m}_{i+1}+ \Delta^{(i)}(X^{j_m}), \Delta^{(p)}(X^q) \right]=\\
&& \back \sum_{r=1}^m  g_{j_1 \dots j_{r-1} s j_{r+1} \dots j_m} C^{sq}_{j_r}  ((\la X^{j_1}_{i+1}+ \Delta^{(i)}(X^{j_1})) \dots
(\la X^{j_m}_{i+1}+ \Delta^{(i)}(X^{j_m}))=0.
\end{eqnarray*}
Eq. (\ref{qlemma}) is central in our analysis; indeed it entails
that
\begin{equation}
\left[ {\rm{Tr}} \left( L_i^{(q)}(\la)^m \right) , {\rm{Tr}} \left(
L_j^{(q)}(\mu)^n \right)  \right] =0 \qquad {\rm{if}} \ i \neq j
\quad \forall m,n, \label{commutation}
\end{equation}
for arbitrary values of the spectral parameters $\la,\mu$. In fact,
assuming $j
>i$, we have
\begin{equation}
 {\rm{Tr}} \left( L_j^{(q)}(\mu)^n \right)=g_{i_1 \dots i_n} (\mu X^{i_1}_{j+1}+ \Delta^{(j)}(X^{i_1})) \dots
(\mu X^{i_n}_{j+1}+ \Delta^{(j)}(X^{i_n})).
\end{equation}
Since $[ {\rm{Tr}} \left( L_i^{(q)}(\la)^m \right),
\Delta^{(j)}(X^{q}))]=0$, and $[ {\rm{Tr}} \left( L_i^{(q)}(\la)^m
\right), X^{q}_{j+1}]=0$, the vanishing commutation relations
(\ref{commutation}) are actually verified.

Let us now consider the quantum (Bending) Hamiltonians as defined via
\begin{equation}\label{Hilm}
\Ha{a}{l}{m}:={\rm{res}}_{\la=0} \frac{1}{\la^{l+1}} {\rm{Tr}} \left( L_a^{(q)}(\la)^m \right),
\end{equation}
where $ a=2,\dots,N, \ l=0,\dots,m$ and $m$ runs in the set of the
exponents of $\alg$. We notice that the set of Hamiltonians
(\ref{Hilm}) contains those defined in \cite{Balrag}, since
${\rm{Tr}}(( \sum_{k=1}^{a} A^q_k )^{m} )= \Ha{a}{0}{m}$; also, they
satisfy the following equality:
\begin{equation}
\sum_{l=0}^m
\Ha{a}{l}{m}=\Ha{a+1}{0}{m}
\qquad a=2,\dots,N \label{coq}
\end{equation}
where, in the case $a=N$ this equation defines $\Ha{N+1}{0}{m}$.

With our definitions, Eq. \ref{qlemma} in particular entails that
Hamiltonians coming from two different Lax matrices of the family
(\ref{eq:2.6}) quantum commute, i.e.,
\begin{equation}
[ \Ha{a}{l}{m}, \Ha{b}{p}{q}]=0
\quad \forall p,q,l,m \qquad {\rm{if}} \  a \neq b\, .
\label{qcoq}
\end{equation}
We are thus left to consider the commutators among Hamiltonians of
the family (\ref{Hilm}) coming from the same Lax matrix. We deem
that using (possibly with suitable modifications) the techniques
discussed in the recent papers \cite{T,CT} based on the notion of
{\em quantum determinant}, the commutativity of such quantities can
be proven in full generality. However, in the last part of the
present paper, we shall show that coproduct methods provide a
complete answer for cubic Hamiltonians (and hence, for
$\alg=su(3)$). Namely, we can show that specific Hamiltonians (or
suitable linear combinations thereof) up to the third order provide
commuting elements; we notice that no quantum corrections to the
straightforward quantisation procedure are required.

To do so, we first show that it holds, still for any value of the
indexes,
\begin{equation}
[\Ha{a}{m}{m} , \Ha{b}{p}{q}]=0 \qquad\text{and}\qquad [
\Ha{a}{0}{m}, \Ha{b}{p}{q}]=0. \label{qcp}
\end{equation}
The first of these identities follows from the fact that
$\Ha{a}{m}{m}={\rm{Tr}} \left( \left(A_{a} \right)^m \right)$, for
$a=2,\dots,N$, are Casimirs elements of $U(\alg)^{\otimes N}$.

Regarding the second of (\ref{qcp}), we know that the Hamiltonians will
commute if $a \neq b$, so we have to consider just the case $a=b$.
In this case we can use  equations (\ref{coq},\ref{qcoq}) to write:
\begin{equation}
[ \Ha{a}{0}{m}, \Ha{a}{p}{q}]=\sum_{l=0}^m [\Ha{a-1}{l}{m},\Ha{a}{p}{q}]=0
\qquad a>2 .
\end{equation}
On the other hand, if $a=2$,
then $ \Ha{2}{0}{m}$ is a Casimir
and (\ref{qcp}) is verified as well.

Our final task is to prove that
\begin{equation}
\left[ {\rm{Tr}} \left( L_a^{(q)}(\la)^m \right) ,
{\rm{Tr}} \left( L_b^{(q)}(\mu)^n \right)  \right] =0 \qquad {\rm{if}}
\quad  m,n=1,2,3, \quad \forall\, a,b. \label{selfcommutation}
\end{equation}
To our end, we use the explicit expressions
\begin{eqnarray}
&& {\rm{Tr}} \left( L_a^{(q)}(\la)^2 \right)= \lambda^2 \Ha{i}{2}{2}
+ \lambda \Ha{a}{1}{2}+ \Ha{a}{0}{2}\\
&& {\rm{Tr}} \left( L_a^{(q)}(\la)^3 \right)= \lambda^3 \Ha{a}{3}{3}
+ \lambda^2 \Ha{a}{2}{3}+ \lambda \Ha{a}{1}{3}+\Ha{a}{0}{3}.
\end{eqnarray}
We want to show that all the seven Hamiltonians appearing here
commute among themselves. Thanks to the result  (\ref{qcp}), we are
left to prove the commutativity of the three quantities
$\Ha{a}{1}{2}, \, \Ha{a}{1}{3},\ \Ha{a}{2}{3}$. We have:
\begin{equation}
[ \Ha{a}{1}{2}, \Ha{a}{1}{3}]=[\Ha{a}{0}{2}+\Ha{a}{1}{2}+\Ha{a}{2}{2},
\Ha{a}{1}{3}]=[\Ha{a+1}{0}{2}, \Ha{i}{1}{3}]=0
\end{equation}
where we used the fact that $[\Ha{a}{0}{2},
\Ha{a}{1}{3}]=[\Ha{a}{2}{2}, \Ha{a}{1}{3}]=0$, equation (\ref{coq})
and equation (\ref{qcoq}).

Since Eqn's $[\Ha{a}{1}{2}, \Ha{a}{2}{3}]=0$ and
$[\Ha{a}{1}{3},\Ha{a}{2}{3}]=0$ can be proven exactly in the same
way, we have thus shown the commutativity of  a complete set of 
the quantised Bending Hamiltonians (\ref{Hilm}) in the case $\alg=su(3)$.

\bigskip {\small {\bf Acknowledgements.} {This work was partially
supported by the ESF project {\em MISGAM} and by the EC--FP6 Marie
Curie RTN {\em ENIGMA} (Contract number MRTN-CT-2004-5652).}
We want to thank O. Ragnisco for discussion about this subject.}

\end {document}